# XPS studies on AlN thin films grown by ion beam sputtering in reactive assistance of $N^+/N_2^+$ ions: Substrate temperature induced compositional variations

Neha Sharma*, S. Ilango, S. Dash and A. K. Tyagi

Material Science Group, Indira Gandhi Centre for Atomic Research

Kalpakkam 603102, TN, India

**Abstract:** We report on an XPS study of AlN thin films grown on Si(100) substrates by ion beam sputter deposition (IBSD) in reactive assistance of $N^+/N_2^+$ ions to unravel the compositional variation of their surface when deposited at different substrate temperatures. The temperature of the substrate was varied as room temperature (RT), 100°C and 500°C. The binding energy of Al-2p, N-1s and O-1s core electrons indicate the formation of 2H polytypoid of AlN. The increase in concentration of AlN with substrate temperature during deposition is elucidated through detailed analysis with calculated elemental atomic concentrations (at. %) of all possible phases at the film surface. Our results show that predominate formation of AlN as high as 74 at. % is achievable using substrate temperature as the only process parameter. This high fraction of AlN in thin film surface composition is remarkable when compared to other growth techniques. Also, the formation of other phases is established based on their elemental concentrations.

## 1. Introduction

Aluminum nitride (AlN) is a wide band-gap semiconductor having potential applications in the fields of electronic and optoelectronic devices [1, 2]. Polycrystalline AlN thin films found their technical importance as optical sensors in UV range, acoustic-optic devices and piezoelectric stress sensors [3-5] . Herein, synthesis of these films is crucial in such applications. In this context, X-ray photoelectron spectroscopy (XPS) is an analytical technique which has been widely used to explore the composition of AlN powders as well as thin films [6, 7]. AlN thin films, grown by various deposition techniques such as direct current (DC) magnetron sputtering [8], ion implantation [9], metal-organic chemical vapor deposition (MOCVD) [10] and pulsed laser deposition (PLD) [11] have been investigated to quantify the fraction of AlN formed as well as impurity elements such as carbon, metallic aluminum and oxygen unintentionally incorporated into the film during deposition process. XPS analysis becomes even more important when thin films are grown by ion beam sputter deposition (IBSD) executed in assistance of reactive flux of gaseous ions. Such non-equilibrium deposition process is often vulnerable towards compositional changes. In case of AlN thin films deposited by reactive assistive IBSD, aluminum atoms are sputtered with a simultaneous supply of reactive flux of nitrogen ($N^+/N_2^+$) which react on the substrate surface to form AlN [12]. Also, an independent control of incident ion energy and current along with the other deposition parameters such as substrate temperature, assisted ion energy as well as current, are expected to influence the formation of AlN. Hence it is always essential to quantify the fraction of AlN that has formed as a function of various deposition parameters. Since reactive formation of AlN takes place on the substrate, one can expect the effect of substrate temperature as an important deposition parameter by varying which compositional presence of AlN and other inadvertently incorporated species can be altered. These compositional variations are shown to affect the material properties of AlN thin films such as optical absorption, thermal conductivity and piezoelectric response [13]. Many studies have been carried out on above mentioned deposition techniques but XPS investigations of AlN thin films grown by reactive assistive IBSD are rather sparse.

Generally in XPS spectrum, information about the gross presence of various compositional constituents is provided by a 1000 eV wide survey scan. Then a ~ 10 eV wide high resolution (HR) spectrum is acquired for each individual elemental peak. Various chemical states of elements can be extracted by a detailed analysis of this HR spectrum from the respective surface core level shifts known as chemical shifts. A monochromatic X-ray source is used to facilitate such analysis by reducing the contribution of Bremsstrahlung continuum and other unwanted satellites. Another important parameter to be noticed is that only 95% of the photoelectrons emerging from the surface of AlN arise from the top 7 nm of the specimen surface. As oxygen has a high affinity towards elemental aluminum, it gets incorporated into AlN during deposition. Also, AlN reacts with atmospheric oxygen as well as humidity present in the environment to form a passivation surface layer at the film-air interface. Thus, in order to obtain information from the original surface of a specimen material, it is either pre-sputtered or as an alternative depth profiling is recommended to facilitate quantitative data collection by eliminating the deleterious effect of surface contamination [14, 15].

In this paper, we are motivated to investigate the effect of substrate temperature on the compositional evolution of AlN thin films deposited by reactive assistive IBSD. For this purpose, thin film samples were prepared at different substrate temperature viz. RT, 100$^{o}$C and 500$^{o}$C. XPS analysis on these thin films was carried out to obtain the information about AlN phase formation and its subsequent quantification only on the surface. In addition to this, entrainment of nitrogen and oxygen in in their various chemical forms like Al-O and N-Al-O are also addressed by quantifying their respective phase fractions.

## 2. Experimental

AlN thin films were grown on Si (100) substrates by IBSD in reactive assistance of nitrogen plasma ($N^+/N_2^+$). The base pressure of the chamber was $3 \times 10^{-7}$ mbar while working pressure was maintained as $4 \times 10^{-4}$ mbar during deposition. Metal atom flux was provided by sputtering the Al-target with an $Ar^+$ ion beam of 500 eV extracted from a 6 cm RF ion source with an ion current of 80 mA. At the same time reactive flux of $N^+/N_2^+$ ions with 90 eV energy and 200 mA ion current was provided directly to the substrate surface by an end-Hall type assisted ion source. Arrangement details of main and assisted ion source in the vacuum chamber is reported in our earlier study [12]. During deposition temperature of the substrate was varied from RT to 100$^{o}$C and then to 500°C. Thickness of the films was measured using a surface profilometer. The topographic features of the films were analysed with scanning electron microscopy (FE-SEM, Supra 55, Zeiss, Germany) at each deposition temperature.

X-ray photoelectron spectroscopy (XPS, M/s Specs, Germany) was used to explore the composition of AlN thin films at each substrate temperature using a monochromatic X-ray source of aluminum with $K_\alpha$ = 1486.6 eV operated at 15 KV and 22 mA. A concentric hemispherical analyzer of 150 mm diameter was used to analyze the photoelectrons with an electron takeoff angle of 90°. The base pressure of the spectrometer was 2.3 x $10^{-10}$ mbar which was maintained to be the same during measurement. The spectrometer was calibrated to the Ag-$3d_{5/2}$ peak at 368.53 eV and C-1s peak at 284.6 eV. All the spectra were recorded with a resolution of 0.25 eV. For binding energy reference, C-1s peak was used.

## 3. Results and Discussion

Figure 1 shows surface morphology of AlN thin films acquired by SEM at different substrate temperatures. Observations of these micrographs imply that all the films are highly dense without any pores. While films prepared at RT and 100°C exhibit similar surface features and appear to be amorphous, samples prepared at 500°C have cluster like features on the surface.

Figure 2(a), 2(b) and 2(c) show the results of XPS survey scans for all the samples deposited at different substrate temperatures. All these survey scans detected only aluminum (Al-2p and Al-2s), nitrogen (N-1s), oxygen (O-1s) and carbon (C-1s) at the surface of each thin film. Little amount of carbon, as indicated by the C-1s peak, is indicative of the mild surface contamination during sample handling. Prominent oxygen peak is appeared at RT as evident by figure 1(a) and reduces as the substrate temperature increases to 100°C and 500°C as shown in figure 1 (b) & 1(c). In order to calculate the elemental atomic concentration (at. %) of all four

elements, high-resolution scans were performed centered around the peaks Al-2p, C-1s, N-1s and O-1s which are called as core level spectra. For this study, intensity vs. binding energy for a particular elemental scan is defined as a peak while one or more components of a peak which are mathematically generated to represent distinct chemical states within the elemental peak are defined as subpeaks. All subpeaks were fitted using a linear combination of Gaussian and Lorentzian shapes, commonly referred as pseudo-Voigt function [16]. Elemental atomic concentration was obtained after calculating the subpeak areas and applying relative sensitivity factors (RSFs) of 0.537, 1.000, 1.800 and 2.930 respectively for Al-2p, C-1s, N-1s and O-1s as recommended by the equipment manufacturer. Deconvolution of each core-level elemental peak and estimation of corresponding compositional variations at each substrate temperature are explained in the following sections.

**3.1 Al-2p peak**

The core-level spectra of peak Al-2p are shown in figure 3 at different substrate temperatures. The nature of different phases formed during deposition can be inferred from the deconvoluted components of these Al-2p peaks. For this, each individual Al-2p peak is deconvoluted using pseudo-Voigt function as one pair of $2p_{3/2}$ and $2p_{1/2}$ spin-orbit split subpeaks. Figure 3(a) shows that at RT, the Al-2p peak is splitted into a higher intensity subpeak Al-$2p_{3/2}$ at 74.0 eV binding energy and a lower intensity subpeak Al-$2p_{1/2}$ appearing at 74.9 eV binding energy position. As the temperature of the substrate was raised to 100°C, Al-$2p_{3/2}$ appeared at the same position with ~74.0 eV binding energy but Al-$2p_{1/2}$ peak occurred at 74.3 eV undergo a larger chemical shift of 0.6 eV on the lower binding energy side as shown in figure 3(b). But no chemical shift was observed for the peak Al-$2p_{3/2}$ as well as for the peak Al-$2p_{1/2}$ which remain at the same positions when the temperature of the substrate is increased to 500°C. These binding energy values are listed in table 1 at each substrate temperature. From these observations, it is clear that each core level Al-2p peak is composed of two types of contributions. One rising from the subpeak Al-2p$_{3/2}$ occurring at ~74 eV which can be assigned to nitidic aluminum in the form of AlN. This AlN retains the original 2H polytypoid (P6$_3$mc) [7, 17-20].

Another contribution originates from the subpeak Al-2p$_{1/2}$ with binding energy lying in the range 74.3 – 74.9 eV and can be attributed to the oxidic aluminum in Al$_2$O$_3$. But the noticeable fact here is that the subpeak Al-2p$_{1/2}$ is associated with a chemical shift of 0.6 eV on the lower binding energy side as the temperature of the substrate is increased from RT to 100$^o$C and remains at the same position when substrate was at 500$^o$C. This chemical shift is a finger print of the valence electrons' bonding environment which in turn affects the overall electrostatic interaction in the atom. Thus changes in the valence environment result in the change in the binding energy of all the inner core electrons [17]. These observations suggest that only a small change in oxidic aluminum bonding environment took place when substrate temperature was raised from RT to 100$^o$C. But no significant change occurs when the temperature of the substrate was raised to 500$^o$C.

**3.2 N-1s peak**

Successive plots of N-1s peak at different substrate temperature and their respective deconvoluted subpeaks are shown in figure 4. Two subpeaks were used in fitting each N-1s peak using pseudo-Voigt line shapes at each substrate temperature. Among them, subpeak with higher intensity is designated as N-1s[#1] and that with lower intensity as N-1s[#2]. Figure 4(a) presents the N-1s peak observed from the AlN thin film grown at RT. The subpeak N-1s[#1] has a binding energy of 396.4 eV while other subpeak N-1s[#2] appeared at 397.6 eV. As the temperature of the substrate is raised to 100$^o$C, no change in the binding energy of subpeaks N-1s[#1] and N-1s[#2] is observed. This is shown in figure 4(b). At 500°C substrate temperature, it can be observed from figure 4(c) that the subpeak N-1s$^{\#1}$ appears at the same position but a significant change is found in the occurrence of the subpeak N-1s$^{\#2}$ which appears at 398.6 eV with a chemical shift of 1 eV on the higher binding energy side. Aforementioned observations establish that the subpeak N-1s$^{\#1}$ has its binding energy ~ 396.4 ± 0.1 eV. This binding energy value is consistent with that nitrogen contribution in N-1s peak which is bound to aluminum in wurtzite hexagonal AlN (i.e. 2H polytypoid). This can be understood by the hypothesis given by Costales et al. [21]. In reactive assistive IBSD, major fraction of the nitrogen flux arriving at the growing AlN film surface is $N^+/N_2^+$ ions. These ions get incorporated into the growing film and progressively increase their coordination to Al-species until a stable single bonded N−Al structure is established in wurtzite hexagonal phase of AlN. At the same time, the subpeak N-1s$^{\#2}$ is observed in the binding energy range of 397.6 eV to 398.6 eV. These binding energy values match well with that contribution of N-1s which forms a three component system Al−O−N, generally referred as aluminum oxynitride. This can be present in spinel phase as well as in amorphous phase. According to L. Rosenberger et al. N-1s$^{\#2}$ is a representative of that nitrogen which is existing in an intermediate electron withdrawing environment where nitrogen is bound to an aluminum, which inturn bound to an oxygen. Thus in the light of the reports existing in literature on XPS studies of aluminum oxynitride, we interpret the emergence of N-1s$^{\#2}$ as $N^+/N_2^+$ ion induced nitriding of $Al_2O_3$ present in the form of tiny domains as well as in the grain boundaries. It is found in very low concentrations as this type of Al−O−N interaction is not energetically favorable all the times. These binding energy values for all the subpeaks of N-1s are listed in table 1.

In general, at. % composition of any compound AB can be calculated using following equation [17]:

$$\frac{n_A}{n_B} = \frac{I_A}{I_B} * \frac{(RSF)_B}{(RSF)_A} \qquad (1)$$

Where $n_A$ and $n_B$ are the at. % concentration of element A and B respectively. $I_A$ and $I_B$ represent their respective area under the XPS peak. To estimate the at. % concentration of AlN in the film at each substrate temperature, Al-$2p_{3/2}$ and N-$1s^{\#1}$ subpeaks were analyzed together. Our calculations suggest that at RT, 46 at. % AlN is formed. At 100°C substrate temperature, interaction of Al with N gets enhanced as compared to RT and 58 at. % of AlN is observed. With further increase in substrate temperature to 500°C, $N^+/N_2^+$ ion flux interacts more efficiently to increase its coordination with Al to form AlN and becomes dominating over other possible elemental interactions. Hence 74 at. % AlN was formed on the surface which is remarkably higher fraction as compared to many other techniques [11, 20]. Thus as a consequence of the increase in substrate temperature, Al-N interaction becomes better, forming larger fraction of AlN. As explained earlier that subpeaks Al-$2p_{1/2}$ and N-$1s^{\#2}$ indicate their respective interaction with oxygen, so high resolution XPS of O-1s is analysed in the next section to delineate the role of oxygen during the growth of AlN thin film by reactive assistive IBSD.

**3.3 O-1s peak**

Successive O-1s spectra from AlN thin films grown at cumulative substrate temperatures are shown in figure 5. The O-1s peak was fitted as two subpeaks using pseudo-Voigt line shapes. The subpeak with higher intensity is designated as O-$1s^{\#1}$ while other with lower intensity is designated as O-$1s^{\#}2$ for all the spectra. One noticeable feature of figure 5 is that, as the temperature of the substrate increases, intensity of O-1s peak decreases indicating lesser incorporation of oxygen at higher substrate temperatures. The subpeak O-$1s^{\#1}$ is found to possess

~531.4 eV binding energy at all substrate temperatures as shown in figures 5(a), 5(b) and 5(c). As suggested by Harris et al. [22], this subpeak can be assigned to oxygen bound to Al with high coordination number forming tiny domains of α-$Al_2O_3$. Thus the subpeak O-$1s^{\#1}$ together with Al-$2p_{1/2}$ is analysed to calculate the fraction of $Al_2O_3$ domains using equation (1). At RT, 31 at. % $Al_2O_3$ was found to form tiny domains. This fractional contribution is decreased to 25.3 at. % as the temperature of the substrate is raised to 100°C. Finally, at 500°C substrate temperature observed fraction of $Al_2O_3$ domains becomes minimum 23.7 at. %.

The subpeak O-$1s^{\#2}$ is obtained with 529.8 ± 0.1 eV binding energy at all substrate temperatures as presented in figures 5(a), 5(b) and 5(c). This subpeak can be attributed to the following three type of contributions. (i). Oxygen bound to aluminum with reduced coordination in the grain boundaries like N−$Al(OH)_2$. (ii). Oxidized aluminum at the grain boundaries of AlN crystals and (iii). O−Al interaction resulting from the reaction of molecular oxygen or water with aluminum at film surface and vacuum interface but with significant contribution from crystal-grain interface [7, 20]. To calculate the percentage fraction of these O−Al interaction at grain boundaries, O-$1s^{\#2}$ together with Al-$2p_{1/2}$ is analysed using equation (1). Total contribution of 10.6 at. % is observed at RT which is reduced to 9.8 at % and 8.3 at. % as the temperature of the substrate is raised to 100°C and 500°C respectively. These values of elemental concentrations are listed in Table. 2. Thus based on above observations and analysis, it is evident that AlN becomes the major constituent of film composition while O− Al interaction also constitutes significant fraction of it. These results are consolidated in figure 6 displaying the compositional variations of these phases at different substrate temperatures.

**Conclusion**

AlN thin films were grown by reactive assistive IBSD on Si(100) substrates at different substrate temperatures viz., RT, 100°C and 500°C. XPS investigations on these samples revealed a clear compositional variation in accordance with substrate temperatures during deposition. The binding energy shift obtained due to various chemical compositions makes it evident that the surface chemical composition is predominately due to wurtzite hexagonal AlN. A mixture of tiny $Al_2O_3$ domains and Al−O interaction due to oxidized aluminum dangling bonds possibly occupy the region between the boundaries of the neighboring AlN grains. For films prepared at RT, 46 at. % AlN was formed along with 31 at. % Alumina domains and 10.6 at. % Al-O bond pairs at the grain boundaries. An increase in concentration of AlN to 58 at. % was observed as the temperature of the substrate was raised to 100°C. Concentration of Al-O bonds from $Al_2O_3$ domains and grain boundaries was 25.3 at. % and 9.8 at. % respectively. A maximum compositional concentration of 74 at. % of AlN was observed on the film surface when the temperature of the substrate was increased to 500°C which is remarkably higher than the reported values by many other techniques. Presence of alumina domains was reduced to a minimum of 23.7 at. % with an associated decrease of oxygen at grain boundaries to 8.3 at. % . It is worth noticing that the increase in substrate temperature helps enhancing the net AlN formation without any change in chemical environment as found from the XPS plots. This is an important aspect wherein substrate temperature could be used effectively as a key parameter to tune the composition of reactive assistive IBSD grown AlN thin films.

## Acknowledgements

The authors would like to thank Dr. G. Amarendra, Director, Materials Science Group for his encouragement and support. The authors would also like to thank Dr. Arindam Das for fruitful discussion, Dr. Shamima Husain and Ms. P. C, Clinsha for XPS measurements and Dr. R. Pandian for SEM.**References**

[1] A.F. Belyanin, L.L. Bouilov, V.V. Zhirnov, A.I. Kamenev, K.A. Kovalskij, B.V. Spitsyn, Application of aluminum nitride films for electronic devices, Diamond and Related Materials, 8 (1999) 369-372.
[2] A. Khan, K. Balakrishnan, T. Katona, Ultraviolet light-emitting diodes based on group three nitrides, Nat Photon, 2 (2008) 77-84.
[3] J.P. Kar, G. Bose, S. Tuli, A study on the interface and bulk charge density of AlN films with sputtering pressure, Vacuum, 81 (2006) 494-498.
[4] K.-H. Chiu, J.-H. Chen, H.-R. Chen, R.-S. Huang, Deposition and characterization of reactive magnetron sputtered aluminum nitride thin films for film bulk acoustic wave resonator, Thin Solid Films, 515 (2007) 4819-4825.
[5] J. Olivares, S. González-Castilla, M. Clement, A. Sanz-Hervás, L. Vergara, J. Sangrador, E. Iborra, Combined assessment of piezoelectric AlN films using X-ray diffraction, infrared absorption and atomic force microscopy, Diamond and Related Materials, 16 (2007) 1421-1424.
[6] J. Binner, Y. Zhang, Surface chemistry and hydrolysis of a hydrophobic-treated aluminium nitride powder, Ceramics International, 31 (2005) 469-474.
[7] L. Rosenberger, R. Baird, E. McCullen, G. Auner, G. Shreve, XPS analysis of aluminum nitride films deposited by plasma source molecular beam epitaxy, Surface and Interface Analysis, 40 (2008) 1254-1261.
[8] A. Mahmood, R. Machorro, S. Muhl, J. Heiras, F.F. Castillón, M.H. Farías, E. Andrade, Optical and surface analysis of DC-reactive sputtered AlN films, Diamond and Related Materials, 12 (2003) 1315-1321.
[9] P.M. Raole, P.D. Prabhawalkar, D.C. Kothari, P.S. Pawar, S.V. Gogawale, XPS studies of N + implanted aluminium, Nuclear Instruments and Methods in Physics Research Section B: Beam Interactions with Materials and Atoms, 23 (1987) 329-336.
[10] H. Liu, D.C. Bertolet, J.W. Rogers Jr, Reactions of trimethylaluminum and ammonia on alumina at 600 K — surface chemical aspects of AlN thin film growth, Surface Science, 340 (1995) 88-100.
[11] N. Laidani, L. Vanzetti, M. Anderle, A. Basillais, C. Boulmer-Leborgne, J. Perriere, Chemical structure of films grown by AlN laser ablation: an X-ray photoelectron spectroscopy study, Surface and Coatings Technology, 122 (1999) 242-246.
[12] N. Sharma, K. Prabakar, S. Ilango, S. Dash, A.K. Tyagi, Application of dynamic scaling theory for growth kinetic studies of AlN-thin films deposited by ion beam sputtering in reactive assistance of nitrogen plasma, Applied Surface Science, 347 (2015) 875-879.
[13] G.A. Slack, L.J. Schowalter, D. Morelli, J.A. Freitas Jr, Some effects of oxygen impurities on AlN and GaN, Journal of Crystal Growth, 246 (2002) 287-298.
[14] S.K. Koh, Y.-B. Son, J.-S. Gam, K.-S. Han, W.K. Choi, H.-J. Jung, Formation of New Surface Layers on Ceramics by Ion Assisted Reaction, Journal of Materials Research, 13 (1998) 2560-2564.
[15] Y. Watanabe, Y. Hara, T. Tokuda, N. Kitazawa, Y. Nakamura, Surface oxidation of aluminium nitride thin films, Surface Engineering, 16 (2000) 211-214.

**Figure Captions:**

**Figure 1:** Morphology of the film surface as seen by using a scanning electron microscope (SEM) at different substrate temperatures.

**Figure 2:** 1000 eV wide survey scans of AlN thin films deposited at (a) RT, (b) 100°C and (c) 500°C substrate temperatures.

**Figure 3:** A series of Al-2p plots deconvoluted in to constituent subpeaks. These plots progress from top to bottom in cumulative substrate temperature as (a) RT, (b) 100°C and (c) 500°C.

**Figure 4:** A series of N-1s plots deconvoluted in to constituent subpeaks. These plots progress from top to bottom in cumulative substrate temperature as (a) RT, (b) 100°C and (c) 500°C.

**Figure 5:** A series of O-1s plots deconvoluted in to constituent subpeaks. These plots progress from top to bottom in cumulative substrate temperature as (a) RT, (b) 100°C and (c) 500°C.

**Figure 6:** Compositional variation of AlN along with $Al_2O_3$ domains and Al-O interaction at the grain boundaries with substrate temperature.

**Tables:**

Table-1: Variation of binding energy of each subpeak with substrate temperature.

| Substrate temperature (°C) | Subpeaks (Binding Energy in eV) | | | | | |
|---|---|---|---|---|---|---|
| | Al-2p$_{3/2}$ | Al-2p$_{1/2}$ | N-1s$^{\#1}$ | N-1s$^{\#2}$ | O-1s$^{\#1}$ | O-1s$^{\#2}$ |
| RT | 74.0 | 74.9 | 396.4 | 397.6 | 531.4 | 529.8 |
| 100 | 74.0 | 74.3 | 396.4 | 397.6 | 531.4 | 529.8 |
| 500 | 74.0 | 74.3 | 396.4 | 398.6 | 531.4 | 529.8 |

Table-2 Atomic concentration at different substrate temperatures

| Substrate Temperature (°C) | Chemical Composition (at. %) | | |
|---|---|---|---|
| | AlN | Al$_2$O$_3$ as tiny domains | Al−O in grain boundaries |
| RT | 46 | 31 | 10.6 |
| 100 | 58 | 25.3 | 9.8 |
| 500 | 74 | 23.7 | 8.3 |

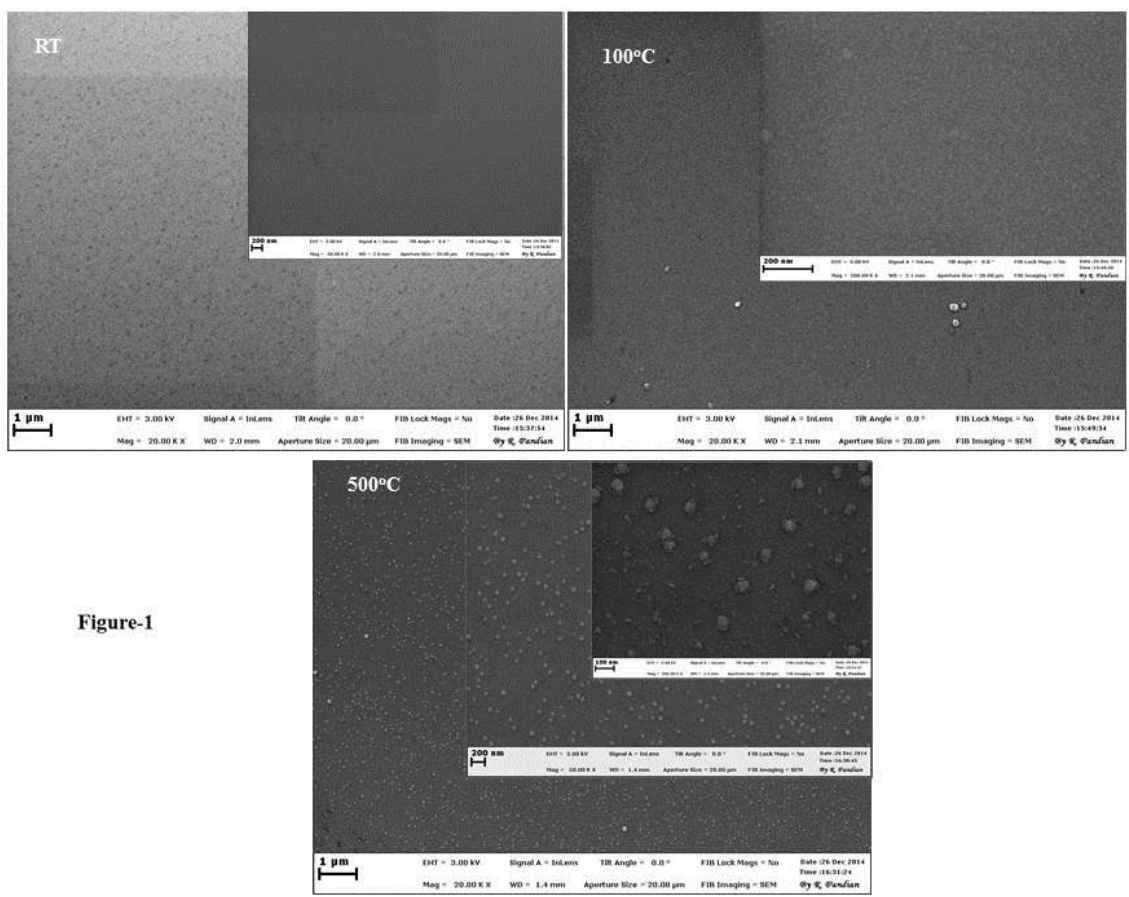

Figure-1

Figure 2

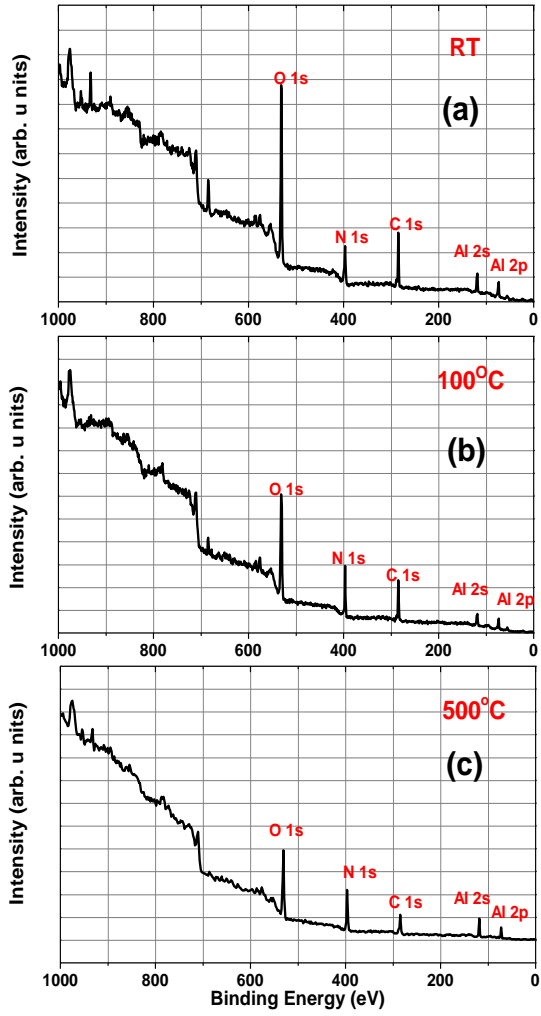

Figure 3

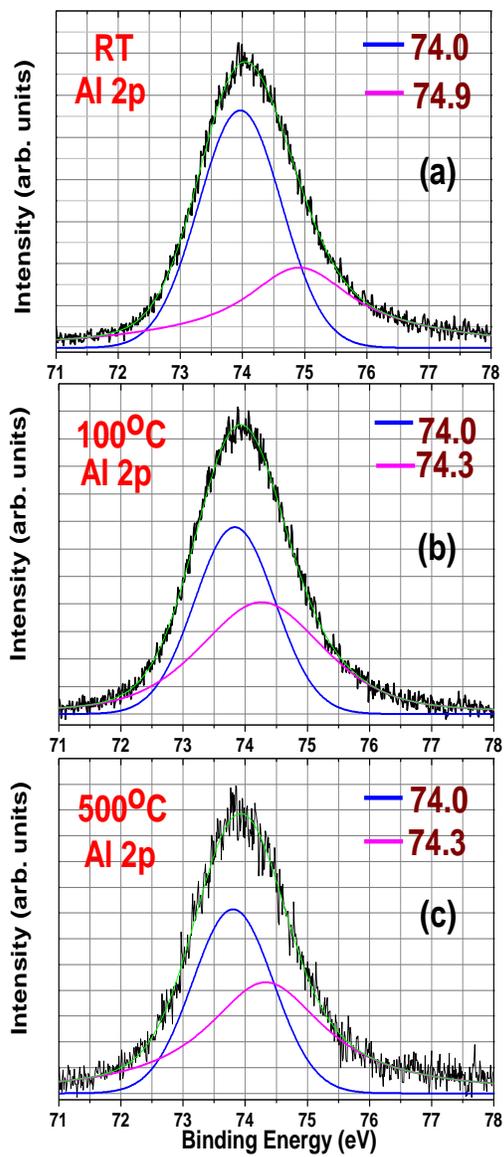

Figure 4

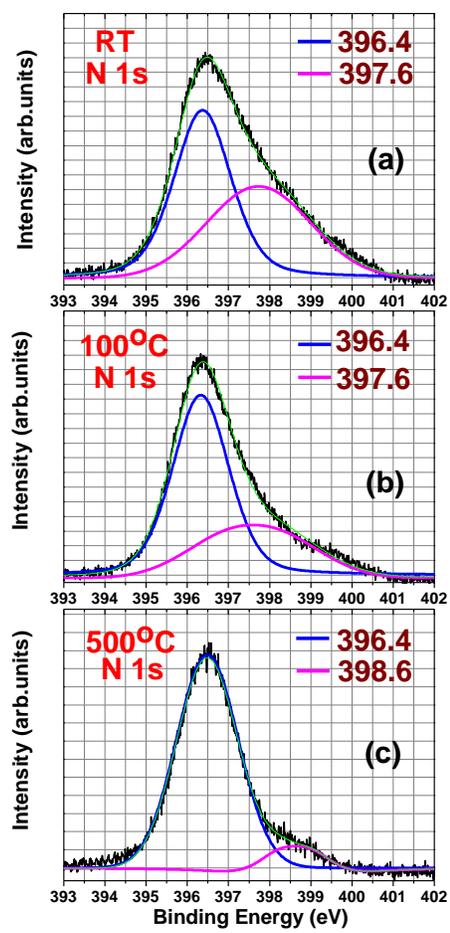

Figure 5

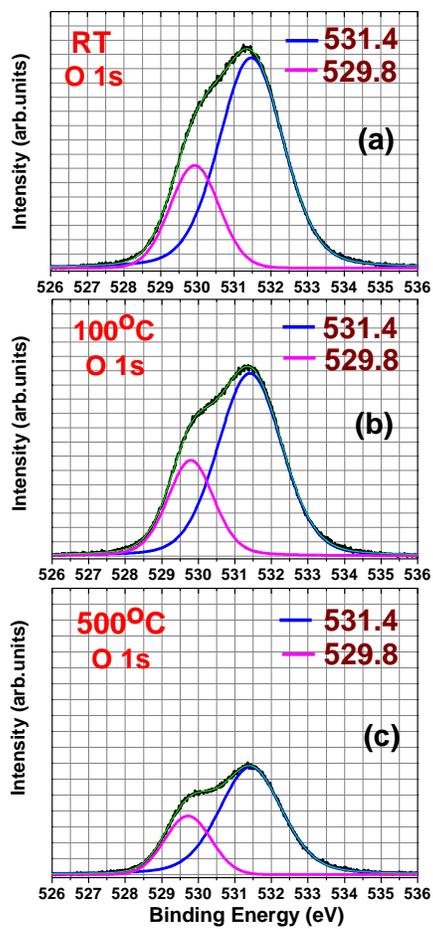

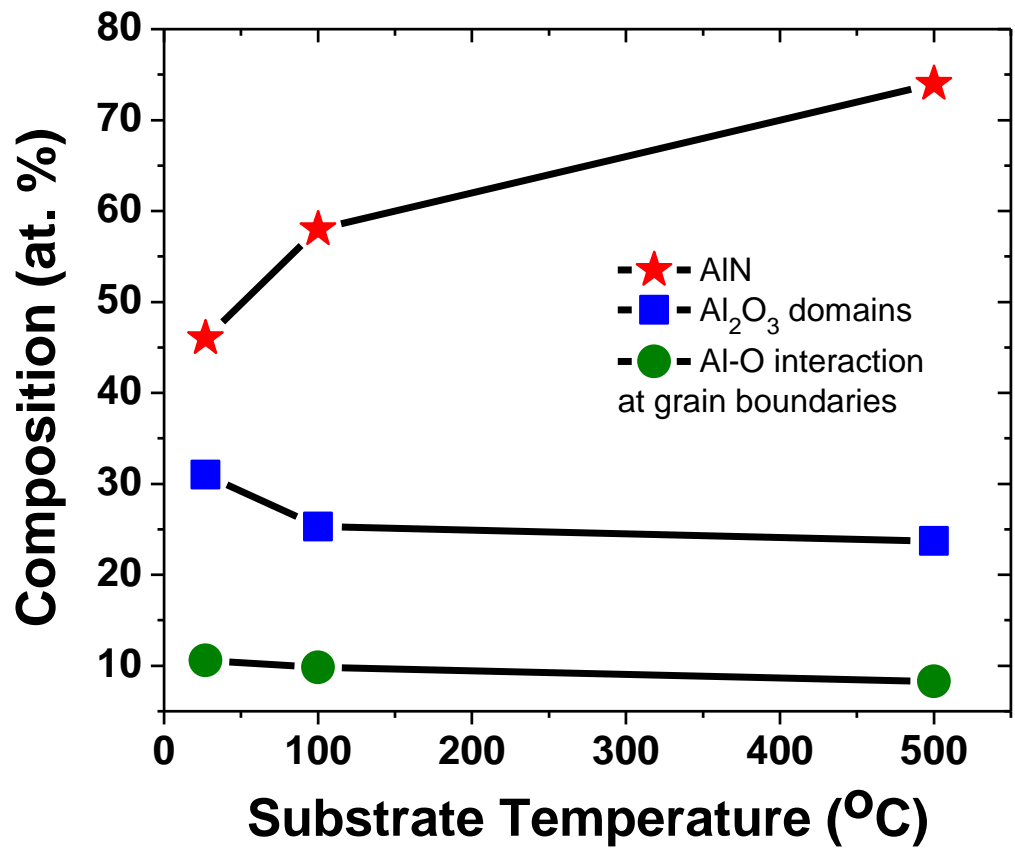

Figure 6